\documentclass[amsmath,amssymb,reprint,prb,superscriptaddress]{revtex4-1}

\usepackage{graphicx} 
\usepackage{dcolumn}  
\usepackage{bm}       

\begin{document}

\title{\textit{Ab initio} computation of $d$--$d$ excitation energies  in
               low-dimensional \\ Ti and V oxychlorides}

\author{Nikolay A.~Bogdanov}
 \affiliation{Institute for Theoretical Solid State Physics, IFW
  Dresden, Helmholtzstr.~20, 01069 Dresden,~Germany} 
 \affiliation {National University of Science and
  Technology MISIS, Leninskiy~pr.~4, 119049 Moscow,~Russia}
\author{Jeroen van den Brink}
\affiliation{Institute for Theoretical Solid State Physics, IFW
  Dresden, Helmholtzstr.~20, 01069 Dresden,~Germany}
\author{Liviu Hozoi}
\affiliation{Institute for Theoretical Solid State Physics, IFW
  Dresden, Helmholtzstr.~20, 01069 Dresden,~Germany}

\date{\today}


\begin{abstract}
Using a quantum chemical cluster-in-solid computational scheme, we calculate the local
$d$--$d$ excitation energies for two strongly correlated Mott insulators, the oxychlorides
TiOCl and VOCl.
TiOCl harbors quasi-one-dimensional spin chains made out of $S\!=\!1/2$ Ti$^{3+}$ ions
while the electronic structure of VOCl displays a more two-dimensional character.
We find in both cases that the lowest-energy $d$--$d$ excitations are within the $t_{2g}$
subshell, starting at $0.34$ eV and indicating that orbital
degeneracies are significantly lifted.
In the vanadium oxychloride, spin triplet to singlet excitations are calculated to be
1 eV higher in energy.
For TiOCl, the computed $d$-level electronic structure and the symmetries of the wavefunctions are
in very good agreement with resonant inelastic x-ray scattering results and optical
absorption data.
For VOCl, future resonant inelastic x-ray scattering experiments will constitute a direct
test of the symmetry and energy of about a dozen of different $d$--$d$ excitations that
we predict here.
\end{abstract}

\maketitle

\section{Introduction}

The vast majority of solid-state electronic structure calculations is based
nowadays on density functional theory (DFT).
For strongly correlated systems, however,
the wavefunction based {\it ab intio} methods certainly provide a viable
alternative.
For this class of materials, of which a typical example are the parent
compounds of the high-$T_{\rm c}$ copper oxide superconductors, it is essential
to properly describe the multiconfigurational character of the many-electron
wavefunction.
The multiconfiguration and multi-reference quantum chemical methods constitute
here a natural choice.
When i) the clusters on which such computational techniques are applied are large
enough and ii) the embedding potential describing the crystalline surroundings
is judiciously constructed, the wavefunction based approach yields results in
excellent agreement with the experimental data.
This has been recently shown for the sequence of spin-state transitions in a
$d^6$ system, LaCoO$_3$,\cite{LaCoO_hozoi_09} and the $d$-orbital electronic 
structure of a number of Cu $d^9$ oxides.\cite{CuO2_dd_hozoi11}  
We used in those studies a fully {\it ab initio} embedding scheme in which the
effective embedding potential is constructed on the basis of prior periodic 
Hartree-Fock (HF) calculations for the extended crystal, in contrast to other
quantum chemical investigations based on simpler point charge
embeddings.\cite{MOCl_note1}
Further, as a general recipe, the embedded cluster that enters the post-HF treatment
comprises in our approach in addition to the ``central'' MO$_6$ octahedron
the nearest-neighbor (NN) octahedra, where M is a transition-metal ion.
The finite charge distribution of the NN octahedra and the tails of ``central''
Wannier orbitals extending to neighboring sites are thus described with high
accuracy. 
For the particular case of $d$--$d$ excitation energies in Cu $d^9$ oxides,\cite{CuO2_dd_hozoi11}
excellent agreement was found with state of the art resonant inelastic x-ray
scattering (RIXS) experiments.\cite{RevModPhys.83.705}
Yet, as compared to earlier quantum chemical calculations on smaller single-octahedron
[CuO$_6$] clusters embedded in arrays of point charges,\cite{PRB.62.702} differences as large as
0.5~eV were observed.\cite{CuO2_dd_hozoi11}

In the present work, we verify the performance of our cluster-in-solid
computational scheme for the case of low-dimensional Ti $d^1$ and V $d^2$ 
Mott insulators, in particular, the TiOCl and VOCl compounds.
The $d$-level electronic structure of TiOCl has been previously addressed by optical
absorption experiments corroborated with model Hamiltonian configuration-interaction
(CI) simulations of the optical spectra,\cite{PRL.95.097203,tmo_rueckamp_05}
DFT-based investigations,\cite{PRB.67.020405,EPL.67.63,PRB.71.153108,JOP.18.10943} 
{\it ab initio} quantum chemical cluster calculations,\cite{TiOCl_Broer_07}
and RIXS measurements.\cite{RIXS_TiOCl}
The V oxychloride compound, on the other hand, is much less investigated.
On the experimental side, basic information concerning its $d$-level electronic
structure is available from optical absorption data.\cite{NJP.10.053027}
Here, we perform a detailed quantum chemical investigation to better understand
the nature 
and symmetry 
of the different excited states identified by optical experiments.
\cite{NJP.10.053027}

\begin{figure}[b]
\includegraphics[angle=0,width=1.0\columnwidth]{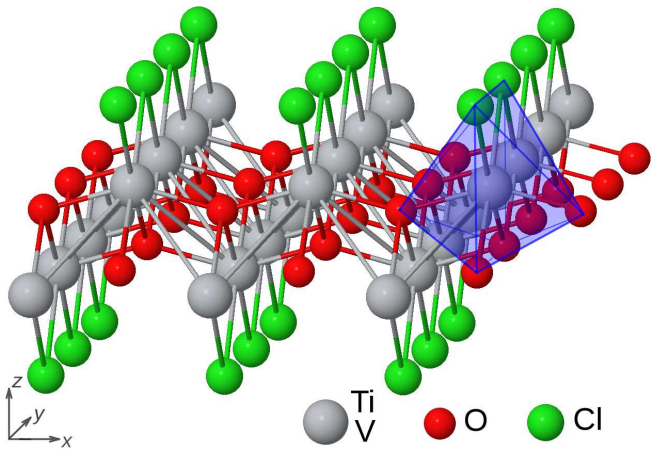}
\caption{Layered crystal structure of TiOCl and VOCl.
A distorted O$_{4}$Cl$_{2}$ octahedron is highlighted on the right hand side.
}
\end{figure}

\section{Structural and computational details}

TiOCl and VOCl display the same highly anisotropic crystalline structure,
with a bilayer network of metal and oxygen sites well separated along the
crystallographic {\it c} axis by double Cl layers (see Fig.\,1).
Each metal ion lies in the center of a distorted O$_4$Cl$_2$ octahedron.
Adjacent octahedra share edges along the {\it b} axis and corners along {\it a}.
The unit cell is orthorhombic, with space group {\it Pmmn}~(number 59).\cite{TiOCl_cryst_snigireva,VOCl_cryst2}

Our computational investigation implies a sequence of several steps.
Periodic restricted HF calculations are first performed with the {\sc crystal}
package.\cite{crystal}
We used the experimental lattice parameters and all-electron Gaussian-type
basis sets. 
Further post-HF correlation calculations are carried out on finite embedded
clusters.
The same type of cluster is employed for both TiOCl and VOCl.
It consists of nine MO$_4$Cl$_2$ octahedra, i.e., one reference octahedron
plus eight neighboring octahedra, where M is either Ti or V.
The reference octahedron in the center of this cluster $\mathcal{C}$
defines the active region $\mathcal{C}_A$ where local $d$--$d$
excitations are explicitly computed. The adjacent eight octahedra
constitute a buffer region $\mathcal{C}_B$. The role of the sites in
$\mathcal{C}_B$ is to provide support for the tails of Wannier
orbitals (WO's) in the $\mathcal{C}_A$ region. 
 
The orbital basis for the correlation calculations is a set of projected
HF Wannier functions. Localized HF WO's are obtained with
the Wannier-Boys localization module \cite{JCP.115.9708} of the {\sc
crystal} package and subsequently projected onto the set of Gaussian basis 
functions associated with the atomic sites of the cluster $\mathcal{C}$.
For our choice of $\mathcal{C}_B$, the norms of the projected WO's centered within
the active region $\mathcal{C}_A$ are not lower than $99.8\%$ of the original
HF WO's.
Additionally, the HF data is used to generate an effective embedding potential for the
nine-octahedra fragment $\mathcal{C}$.
This potential is obtained from the Fock operator in the restricted HF calculation
and models the surroundings of the finite cluster, i.e., the
remaining of the crystalline lattice.

While the doubly occupied WO's in the buffer region $\mathcal{C}_B$ are kept frozen, all valence orbitals
centered at ligand and metal sites in $\mathcal{C}_A$ (and their tails in $\mathcal{C}_B$) are
further re-optimized in multiconfiguration complete-active-space self-consistent-field
(CASSCF) calculations. 
In the latter, the ground-state wavefunction and the crystal-field excited states at
the central Ti/V site are computed in state-averaged multiroot calculations.
The $d$-level splittings are finally obtained by additional
single and double CI (MRCI) calculation. \cite{[{For a monograph, see }][{}]Helgaker} 
The whole procedure is described in more detail in Ref.~\onlinecite{CuO2_dd_hozoi11}.

The CASSCF and MRCI investigations are carried out with the {\sc molpro} package.
\cite{molpro_2006}
The effective embedding potential is added to the one-electron Hamiltonian in the
CASSCF/MRCI computations with the help of the {\sc crystal-molpro} interface.
\cite{crystal_molpro_int}

We used O and Cl basis sets of triple-zeta quality as designed by Towler {\it et al.}
 \cite{Towler_O_basis95} and Prencipe {\it et al.} \cite{Prencipe_Cl_basis95}
For the Ti$^{3+}$ and V$^{3+}$ species, we employed the triple-zeta $s$ and $p$
basis sets of Mackrodt {\it et al.} \cite{Mackrodt_Tisp_basis97} and Dovesi
{\it et al.} \cite{Dovesi_Vsp_basis97}, respectively, plus triple-zeta $d$ basis 
functions as designed by Ahlrichs {\it et al.} \cite{Ahlrichs_basis92}

\section{Results and discussion}

\subsection {TiOCl}

The magnetic properties of TiOCl have been the topic of extensive investigations
over the last decade.
The largest intersite interactions are antiferromagnetic (AF), occur between
NN Ti$^{3+}$ $S\!=\!1/2$ sites along the $b$ axis, and give rise to AF spin chains 
along that direction.
The interchain couplings are also sizable and responsible for geometrical
frustration.

The doubling of the unit cell in the low-temperature nonmagnetic phase
\cite{TiOCl_XRD_05,TiOBr_XRD_06} suggests a spin-Peierls dimerized
ground state at low temperatures.
The observed deviations from conventional spin-Peierls behavior, e.g.,
the existence of two distinct phase transitions at 91 and 67~K, were
attributed to the above mentioned  frustration of the interchain
interactions.\cite{PRL.95.097203}

\begin{table}[t]
\caption{Ti$^{3+}$ $d$-level splittings in TiOCl.
A non-standard local reference system is chosen for the octahedrally coordinated
Ti ion, see Fig.~1, with a rotation of $45^{\circ}$ around $x$.
The ground-state $d_{y^2-z^2}^1$ configuration is taken as reference.
All energies are given in eV.}
\begin{ruledtabular}
\begin{tabular}{lcc}
Orbital occupation &CASSCF/MRCI   &CASSCF/CASPT2\footnotemark[1] \\
\colrule
\\
$d^1_{y^2-z^2}$        &0/0           &0/0       \\
$d^1_{xy}$             &0.36/0.34     &0.29/0.29 \\
$d^1_{xz}$             &0.65/0.63     &0.66/0.68 \\
$d^1_{yz}$             &1.51/1.55     &1.59/1.68 \\
$d^1_{x^2}$            &2.35/2.36     &2.30/2.37 \\
\end{tabular}
\footnotetext[1]{Ref.~\onlinecite{TiOCl_Broer_07}.}
\end{ruledtabular}
\end{table}

As concerns the electronic structure of TiOCl,
DFT calculations within the local density approximation
(LDA) indicate sizable occupation numbers for each of the Ti
$t_{2g}$ levels.
\cite{PRB.67.020405,EPL.67.63,PRB.71.153108,JOP.18.10943}
Early interpretations of the unusual physical properties of this
material thus invoked the presence of strong orbital fluctuations.
\cite{PRB.67.020405,EPL.67.63,PRB.70.134429}
Inclusion of local Mott-Hubbard type correlations within dynamical
mean-field theory (DMFT) and LDA+DMFT calculations effectively pulls
one of the $t_{2g}$ levels below the other two components and reduces
the occupation of the latter to nearly~0.
\cite{PRB.71.153108,JOP.18.10943}
On the theoretical side, this is confirmed by earlier quantum chemical
cluster calculations \cite{TiOCl_Broer_07} as well by our present results.

\begin{table}[b]
\caption{Ti $d$--$d$ excitation energies in TiOCl, quantum chemical vs.~experimental
results. 
A $J_b\ln2$ term was here substracted from each of the RIXS values reported in
Ref.~\onlinecite{RIXS_TiOCl}, see text.
All energies are given in eV.}
\begin{ruledtabular}
\begin{tabular}{llll}
Orbital occupation   &MRCI         &RIXS\footnotemark[1] &Optics\footnotemark[2]\\
\colrule
\\
$d^1_{y^2-z^2}$      &$0$          &$0$      &$0$   \\
$d^1_{xy}$          &$0.34$       &$0.32$   &$-$   \\
$d^1_{xz}$          &$0.63$       &$0.55$   &$0.65$\\
$d^1_{yz}$          &$1.55$       &$1.41$   &$1.50$\\
$d^1_{x^2}$         &$2.36$       &$2.01$   &$-$   \\
\end{tabular}
\footnotetext[1]{Ref.~\onlinecite{RIXS_TiOCl}.}
\footnotetext[2]{Ref.~\onlinecite{tmo_rueckamp_05}.}
\end{ruledtabular}
\end{table}

In Table I, CASSCF and MRCI results for TiOCl are listed.
The active orbital set contains all five $3d$ functions at the central Ti
site and the singly occupied $3d_{y^2-z^2}$ orbitals at the eight Ti NN
sites.
We considered high-spin intersite couplings, i.e., a ferromagnetic (FM) configuration
of Ti $d$ spins.
Although the WO's at the atomic sites of $\mathcal{C}$ are derived from
periodic restricted HF calculations for the Ti $3d^1$ electron configuration,
the embedding potential is obtained by replacing the Ti$^{2+}$ $3d^1$
ions by closed-shell V$^{2+}$ $3d^{2}$ species.
This is a  good approximation for the farther $3d$-metal sites, as the
comparison between our results and RIXS data shows.
An extension of our embedding scheme toward the construction of open-shell
embeddings is planned for the near future.
In addition to the active Ti $3d$ functions in $\mathcal{C}_A$ and $\mathcal{C}_B$,
all O $2p$ and Cl $3p$ orbitals at the central octahedron are included in the
CISD calculations.

As expected for a $d^1$ system, the MRCI treatment brings minor corrections 
to the $d$--$d$ CASSCF excitation energies. 
Our results are compared in Table I with earlier CASSCF and CASPT2 (complete-active-space
second-order perturbation theory) data from Ref.~\onlinecite{TiOCl_Broer_07}.
A major difference between the present and earlier quantum chemical calculations
is the size of the embedded clusters, i.e., nine octahedra here and one
octahedron in Ref.~\onlinecite{TiOCl_Broer_07}. 
In contrast to other systems, e.g., layered Cu $d^9$ oxides such as La$_2$CuO$_4$
and Sr$_2$CuO$_2$Cl$_2$ where differences of about 0.5~eV were found as function
of the cluster size (see the discussion in Ref.~\onlinecite{CuO2_dd_hozoi11}),
the agreement between the two sets of results is good for TiOCl.

\begin{table*}[ht]
\caption{V $d$--$d$ excitation energies in VOCl.
A non-standard local reference system is chosen, see Fig.\,1,
with a rotation of $45^{\circ}$ around $x$.
The dominant ground-state configuration is $d_{y^2-z^2}^{1}d_{xy}^1$, with a
weight of 81$\%$.
For each of the excited states, the weight of the dominant configuration is
not smaller than 64$\%$.
All energies  are given in eV.
}
\begin{ruledtabular}
\begin{tabular}{llllll}
\multicolumn{2}{l}{Dominant configuration}   &CASSCF\footnotemark[1]
                                                     &CASSCF\footnotemark[2]
                                                             &MRCI\footnotemark[2]
                                                                     &Optics\footnotemark[3] \\
\colrule
\\
$t_{2g}^{2}$, $S\!=\!1$
                   &$d^1_{y^2-z^2}d^1_{xy}$  &0     &0       &0     &0   \\
                   &$d^1_{y^2-z^2}d^1_{xz}$  &0.36  &0.36    &0.34  &0.3 \\
                   &$d^1_{xy}d^1_{xz}$       &0.43  &0.44    &0.46  &0.4 \\
\\
$t_{2g}^{1}e_g^{1}$, S\!=\!1
                   &$d^1_{xy}d^1_{yz}$       &1.61  &1.60    &1.73  &1.5--1.9 \\
                   &$d^1_{y^2-z^2}d^1_{x^2}$ &1.74  &1.71    &1.82  &1.5--1.9 \\
                   &$d^1_{xz}d^1_{yz}$       &1.75  &1.73    &1.88  &1.5--1.9 \\
                   &$d^1_{y^2-z^2}d^1_{yz}$  &2.85  &2.82    &2.55  & \\
                   &$d^1_{xz}d^1_{x^2}$      &3.53  &3.52    &3.17  & \\
                   &$d^1_{xy}d^1_{x^2}$      &3.53  &3.53    &3.19  & \\
\\
$t_{2g}^{2}$, $S\!=\!0$
                   &$d^2_{y^2-z^2}$          &      &1.27    &1.47  &1.1--1.4 \\
                   &$d^1_{y^2-z^2}d^1_{xy}$  &      &1.28    &1.47  &1.1--1.4 \\
                   &$d^1_{y^2-z^2}d^1_{xz}$  &      &1.56    &1.75  &1.5--1.9 \\
                   &$d^2_{xz}$               &      &1.79    &1.97  &1.5--1.9 \\
                   &$d^1_{xy}d^1_{xz}$       &      &1.81    &2.00  &1.5--1.9 \\
\end{tabular}
\footnotetext[1]{High-spin    V $d^1_{y^2-z^2}d^1_{xy}$ NN's, see text.} 
\footnotetext[2]{Closed-shell V $d^2_{y^2-z^2}$         NN's.}
\footnotetext[3]{Ref.~\onlinecite{NJP.10.053027}.}
\end{ruledtabular}
\end{table*}

Good agreement is also found with RIXS and optical absorption experimental 
data, with deviations not larger than 0.1 eV.
The only exception is the $d_{y^2-z^2}$ to $d_{x^2}$ excitation, which is
predicted to occur at 2.36~eV in the quantum chemical calculations and
observed at about 2~eV in RIXS, see Table II.
In the optical absorption measurements,\cite{PRL.95.097203,tmo_rueckamp_05} this
particular transition as well as the $d_{y^2-z^2}$ to $d_{xy}$ excitation were not
identified, which is related to the different selection rules for RIXS and optical absorption.

We recall at this point that in a first approximation the total energy of a 
given state within the $d^n$ manifold is a sum of a crystal-field contribution,
i.e., an on-site crystal-field splitting, and a magnetic term
(see also the discussion in Refs.~\onlinecite{RevModPhys.83.705,CuO2_dd_hozoi11}).
For the ground-state configuration of TiOCl, the NN spin interactions along Ti
chains parallel to the $b$ axis are quite large, $J_b\!\approx\!57$ meV,
\cite{PRB.67.020405} and AF.
From the exact Bethe-ansatz solution of the one-dimensional Heisenberg Hamiltonian,
\cite{Hulthen} the AF ground-state stabilization energy is $J\ln2$.
On the other hand, from overlap considerations, the (super)exchange with the
NN Ti $d_{y^2-z^2}$ spins is either zero or much weaker for the crystal-field
excited states.
Since the quantum chemical calculations were performed for a FM cluster,
\cite{MOCl_note2}
for a meaningful comparison between the MRCI and RIXS data, we subtracted in Table II
from the relative RIXS energies reported in Ref.~\onlinecite{RIXS_TiOCl} a term
$J_b\ln2\!\approx\!0.04$~eV
representing the energy stabilization of the AF ground-state with respect
to the crystal-field excited states.

\subsection{VOCl}

In the family of correlated transition-metal oxide materials, the vanadium
compounds display an impressive variety of vanadium valence states,
crystalline structures, and physical properties.
The valence configuration, for example, may vary from V$^{2+}$ $d^3$
in VO$_{1-x}$ \cite{PRB.5.2775, PRB.68.220403} to V$^{4+}$ $d^1$ 
in CaV$_2$O$_5$ \cite{JCP.120.961} and NaV$_2$O$_5$.\cite{PRL.89.076407}
While the latter vanadates display a layered lattice with ladders of
O-bridged V sites, VO$_{1-x}$ has a rocksalt crystalline structure.
At the level of {\it ab initio} wavefunction calculations, the V oxides are
rather unexplored.
To our knowledge, only the ladder vanadates have been investigated by advanced
quantum calculations.\cite{JCP.120.961, PRL.89.076407, PRL.88.056405}
 
Besides having two electrons instead of one within the $t_{2g}$ set of orbitals,
VOCl also displays in comparison to TiOCl a more pronounced two-dimensional
character as concerns the electronic structure and magnetic
properties.\cite{JPCssp.16.5339}
In Table III, V $d$--$d$ excitation energies for the VOCl compound are given.
Two different sets of CASSCF calculations were carried out, for two different
electron configurations at each of the V NN's.
Preliminary CASSCF calculations were first performed with eighteen orbitals in the
active space, i.e., two $t_{2g}$ orbitals at each of the V sites in the nine-octahedra
cluster.
As for TiOCl, we considered high-spin intersite couplings, i.e., a FM cluster.
It turns out that the dominant ground-state configuration at a given V site is
$d^1_{y^2-z^2}d^1_{xy}$.
The crystal-field excited states for the central V ion are then obtained in 
state-averaged multiroot computations where a group of three additional orbitals was
added to the active space.
To be sure that the three additional orbitals are localized at the central V site
and to allow only for on-site $d$--$d$ excitations, the occupation of the NN
$d_{y^2-z^2}$ and $d_{xy}$ functions was restricted to 1.
Crystal-field splittings obtained from such SCF calculations are listed in the
third column of Table III.
To avoid complications related to states that display low-spin intersite couplings,
only the high-spin on-site excitations were computed in this case.

We note that the ground-state electron configuration in the periodic restricted HF
calculation is $d^1_{y^2-z^2}d^1_{xy}$ but in the construction of the HF embedding
potential we imposed a closed-shell V $d^2_{y^2-z^2}$ configuration for the
remaining part of the lattice.
In a second set of CASSCF calculations, we imposed a closed-shell $d^2_{y^2-z^2}$
electron configuration for the V NN's in the $\mathcal{C}_B$ region of the cluster
as well.
Minor variations were found as compared to the CASSCF $d$--$d$ excitation energies
with high-spin $d^1_{y^2-z^2}d^1_{xy}$ NN's, i.e., less than 0.01 eV for the
lowest three roots and less than 0.03 eV for the higher roots.
The CASSCF V crystal-field splittings for closed-shell V NN's are given
in the fourth column of Table III.
The latter CASSCF wavefunctions were used as reference wavefunctions in further
MRCI calculations. Those MRCI results are listed in the fifth column of Table III.
Only the V $3d$, O $2p$, and Cl $3p$ orbitals at the central octahedron were
correlated in the MRCI treatment.

The agreement between our quantum chemical data and optical absorption results 
reported in Ref.~\onlinecite{NJP.10.053027} is 
very 
good, especially for the lowest
two excitations (see Table III).
The small difference between those excitation energies indicates that the 
splitting between the $d_{y^2-z^2}$ and $d_{xy}$ levels, singly occupied in
the ground-state configuration, is also small, about 0.1 eV.
On the other hand, the energy separation between the $d_{xy}$ and $d_{xz}$ levels
is substantially larger, 0.3 eV, and indicates that orbital fluctuations 
are unlikely to be important in VOCl.
As concerns the $S\!=\!0$ singlet excitations, they acquire very low intensity in the
optical spectra.
On the experimental side, a better characterization of those states call for
high-resolution RIXS measurements.

\section{Conclusions}
With a quantum chemical cluster-in-solid computational scheme, we have determined
the $d$--$d$ excitation energies of TiOCl and VOCl.
The ground state configuration is $t_{2g}^1$ for Ti and high-spin $t_{2g}^2$ for V.
The lowest-energy $d$--$d$ excitations are for both materials within the $t_{2g}$
subshell, starting at $0.34$ eV.
We conclude that therefore orbital degeneracies are lifted to a similar and significant
extent in the two systems, which excludes the presence of strong orbital fluctuations
in the ground state.
In the vanadium oxychloride, spin triplet to singlet excitations start at an excitation
energy of 1.47 eV and above.
The computed $d$-level electronic structure and the symmetries of the wavefunctions are
in very good agreement with RIXS results and optical absorption data for TiOCl.
For VOCl, future RIXS experiments will constitute a direct and stringent test of the
symmetry and energy of about a dozen of different $d$--$d$ excitations that we predict
here.

\begin{acknowledgments}
We thank L.~Siurakshina, J.~Geck, R.~Claessen, and M.~Gr\"{u}ninger for useful discussions.
N.~B. acknowledges financial support from the Erasmus Mundus Programme of the European Union.
L.~H. acknowledges financial support from the German Research Foundation
(Deutsche Forschungsgemeinschaft,
DFG).
\end{acknowledgments}

%






\begin{thebibliography}{48}
\bibitem [{MOC({\natexlab{b}})}]{MOCl_note2}%
  \BibitemOpen
  \bibinfo {note} {The virtual orbital space in the MRCI calculations cannot be
  presently restricted just to the $\mathcal{C}_A$ region. It thus includes
  virtual orbitals in both $\mathcal{C}_A$ and $\mathcal{C}_B$, which leads to
  very large MRCI expansions, $\sim\!10^{10}$ Slater determinants for the
  nine-octahedra FM cluster. For practical reasons, we thus restrict the MRCI
  calculations to FM allignment of the Ti $d$ spins.}\BibitemShut {Stop}%
\end{thebibliography}

\providecommand{\noopsort}[1]{}\providecommand{\singleletter}[1]{#1}%

\end{document}